# Sensorless Resonance Tracking of Resonant Electromagnetic Actuator through Back-EMF Estimation for Mobile Devices

Youngjun Cho
UCL Interaction Centre, University College London, London, United Kingdom
* *y.cho@cs.ucl.ac.uk*

**Abstract**
Resonant electromagnetic actuators have been broadly used as vibration motors for mobile devices given their ability of generating relatively fast, strong, and controllable vibration force at a given resonant frequency. Mechanism of the actuators that is based on mechanical resonance, however, limits their use to a situation where their resonant frequencies are known and unshifted. In reality, there are many factors that alter the resonant frequency: for example, manufacturing tolerances, worn mechanical components such as a spring, nonlinearity in association with different input voltage levels. Here, we describe a sensorless resonance tracking method that actuates the motor and automatically detects its unknown damped natural frequency through the estimation of back electromotive force (EMF) and inner mass movements. We demonstrate the tracking performance of the proposed method through a series of experiments. This approach has the potential to control residual vibrations and then improve vibrotactile feedback, which can potentially be used for human-computer interaction, cognitive and affective neuroscience research.

## I. Introduction
While the visual display of consumer electronics such as smartphones has become one of the most fundamental and effective means in connecting a person to graphical contents and virtual worlds, mechanical keypads in such devices have been disappearing, allowing the screen to be larger and an embedded vibration motor to instead provide vibration feedback. Although there are individual differences in preferences for vibro-tactile effects, having a better vibration actuator and its feedback has been regarded as an important factor in designing state-of-the-art mobile devices and user experiences [1–6].

Various mechanisms have been proposed to make vibration actuators capable of effectively stimulating Pacinian Corpuscles, one of human mechanoreceptors, which responds to rapid vibration of low frequency (up to approximately 500 Hz) [7,8], through a mobile device. Along with the type of mechanisms, actuators can generally be categorized into three groups [4]: i) linear electromagnetic actuators which produce linear actuations enabled by a one-degree-of-freedom mechanical oscillator, ii) rotary electromagnetic actuators that convert direct current (DC) into rotary force, and iii) non-electromagnetic actuators based on the use of smart materials (e.g., piezoelectric materials [9,10], electro-active polymers [11]). In the consumer electronics industry, the linear and rotary electromagnetic actuators have been dominantly used because of their better mechanical durability and inexpensive cost than those of the latter. In particular, the resonant electromagnetic actuator and eccentric rotating mass motor, which are special types of the linear electromagnetic actuators and the rotary electromagnetic actuators, respectively [4], are known as the most commercially successful vibration motors, given that they consume relatively low electrical power, but are still capable of generating sufficient vibration force [11,12]. By contrast with the eccentric rotating mass motors in which the vibration force is coupled with the frequency element [11,13], the force of the resonant electromagnetic actuator is controllable at a specific range of frequency. This is enabled by its mechanism that employs the mechanical linear resonance phenomenon to maximize its vibration force at a given frequency and limited driving energy. This also helps the resonant electromagnetic actuator, which is also commercially branded *Linear Resonant Actuator* (*LRA)*, to respond faster than the rotary motors. Hence, the actuator has been employed more and more recently in mobile devices.

To take an advantage from the mechanical resonance on which a resonant electromagnetic actuator relies, its resonant frequency needs to be either known prior to the use or automatically detectable. Although most manufacturers of the actuator set a specific natural frequency and provide the information, the frequency is often shifted due to many external and internal factors. For instance, driving the actuator for a long period of time or dropping it tear down its internal components such as a mechanical spring, leading to changes of its stiffness and then natural frequency. In addition, nonlinearity in the mechanical mass-spring-damper system exists in reality given many reasons such as the heat production, also leading to shifts of its resonance [14]. The mismatch between frequencies of electrical driving signals and the resonance induces a drop in the vibration force and difficulty in controlling residual vibrations [15]. Thus, an automatic resonance tracking approach is required to address such issue.

A vibration is produced by the movement of the inner mass, which can be tracked through the use of additional sensors such as the hall effect sensors [16] and piezoelectric material [10]. In the real world, however, it is difficult to add an additional sensing channel to inner compact spaces of a mobile device and of an actuator. Hence, we focus on the automatic resonance tracking without the use of a physical sensor. This can be achieved by the estimation of back electromotive force, so-called, *back EMF*, which is the voltage induced by the magnetic linkage flux variation in accordance with Lenz's law, as used in other mechanical actuating systems for controlling torque and displacements [17–20]. For this, we first analyze the actuation model and build a new sensorless resonance tracking algorithm that drives an actuator, makes an inner coil to be in high impedance, and then estimates the back EMF so as to automatically detect the damped (unknown or shifted) natural frequency. This can be of help in keeping the vibration performance from being deteriorated and in extending lifetimes of actuators. The proposed drive scheme produces further potential benefits in designing vibrotactile effects. In particular, a residual vibration, one of the challenges in controlling a mechanical oscillation system [15,21], can be controlled through the estimation of back EMF, potentially helping to improve a person's vibrotactile perception.

This paper is organized as follows: first, we introduce details of commercialized resonant electromagnetic actuators which have been widely used in mobile devices in Section II. Section III discusses the mechanical actuation model commonly used for the actuators. Next, we analyze the electro-magnetic circuit combined with the mechanical structure to derive the relationship between the natural frequency and the back electromotive force, and propose the sensorless resonance tracking algorithm in Section IV. Section V describes conducted experiments and results. Finally, we conclude the paper in Section VI.

## II. Resonant Electromagnetic Actuators for Mobile Devices

A resonant electromagnetic actuator is a typical type of linear electromagnetic actuators developed for handheld consumer electronics mobile devices to convey vibration feedback. The schematic mechanical structure of the actuator is illustrated in Fig.1. This is on the basis of the linear oscillation mechanism. Based on this structure, two types are commercially available in accordance with different coil compositions: the first type is a voice coil based actuator (i.e., coil without ferromagnetic component), such as, *Linear vibration motor (also called Linear resonant actuator)* [22], the other is a solenoid based one (i.e., coil with ferromagnetic materials), such as [23]. By reflecting the sensitive frequency range of the human mechanoreceptor [8,24], the resonant frequency of both types is generally set to a value between 150Hz to 250Hz.

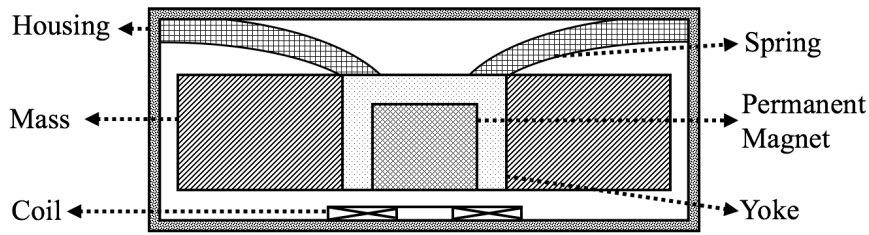

**Fig. 1.** Schematic mechanical structure of the resonant electromagnetic actuator.

Both types of actuators tend to have a high Q factor, which draws a high resonance sharpness [25], so as to maximize their vibration force at a given natural frequency. However, the high Q value, at the same time, narrows their operable frequency range. As discussed above, a shifted damped natural frequency dramatically decreases the vibration force of an actuator with the high Q value when it is driven at a preset frequency provided by its manufacturer. For example, the natural frequency can be easily shifted along with different peak voltage levels of input driving signals as shown in Fig. 2. Here, we tested a commercial resonant electromagnetic actuator (LRA, SEMCO 1036) whose resonant frequency is designed to be 175Hz. The frequency responses collected through the use of an accelerometer and a signal generator indicate the higher the peak voltage of the input signal, the lower the actual resonant frequency (e.g., 175.5Hz at 1.0 V$_{peak}$ and 173.4Hz at 1.5 V$_{peak}$ from this pilot test). Therefore, it is required to automatically track the varied resonant frequency when using the actuator.

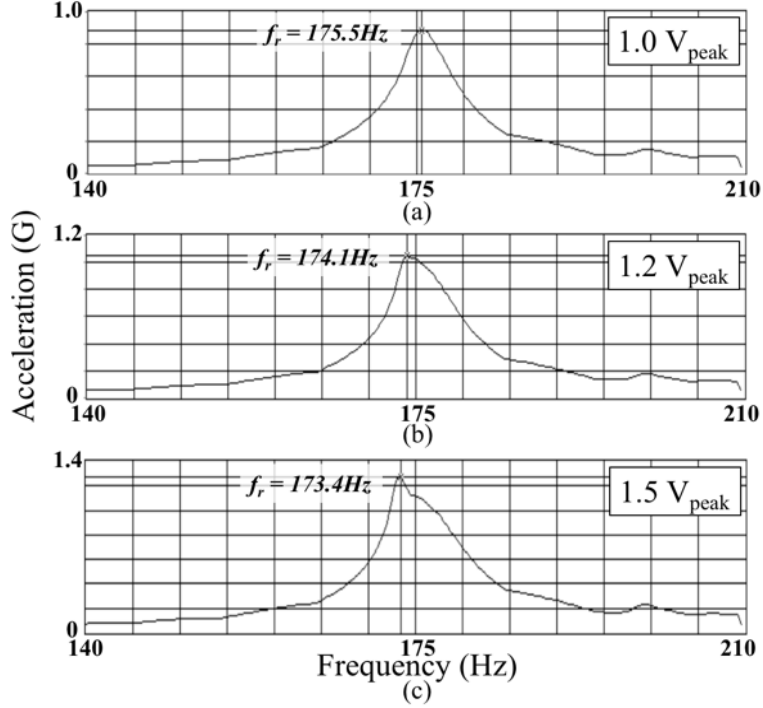

**Fig. 2.** Measured the shifted resonant frequency of the resonant electromagnetic actuator (Product: LRA – SEMCO 1036) along with the different peak voltage of input signal on the frequency response: (a) 175.5Hz - 1.0V$_{peak}$, (b)174.1Hz - 1.2V$_{peak}$, (c)173.4Hz - 1.5V$_{peak}$.

## III. Actuator Model

Fig. 3 shows the electrical-mechanical model of the resonant electromagnetic actuator mounted in the mobile device. The combined system consists of an electrical circuit of a single coil and an equivalent mechanical model of its surroundings. The external excitation force relies on the electrodynamic force generated by the magnetic flux between the coil and the permanent magnet. Given the one degree of freedom structure, we can assume that a direction of current flow inside the coil is vertical to the magnetic flux linkage and the electrodynamic force can be estimated by Fleming's rule. Thus, the external excitation force can be controlled by the electrical input $u$:

$$F(t) = K_f u(t) \qquad (1)$$

where $u(t)$ is the time-varying input signal, $K_f$ is the force constant. In consideration of the operating principle of the actuator that employs a harmonic voltage source, Eq (1) can be rewritten as

$$F(t) = K_f V_A \sin(\omega t) \qquad (2)$$

where $V_A$ is the peak voltage of the input signal, $\omega$ is a frequency of the applied harmonic input source.

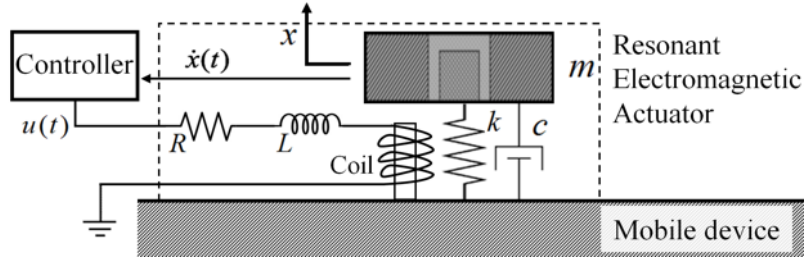

**Fig. 3.** Combined electrical circuit and mechanical model of the actuator embedded in mobile devices.

In the Laplace domain, the dynamic equation of the linear model illustrated in Fig. 3 can be expressed with initial conditions (i.e., initial displacement $x(0)$, initial velocity $\dot{x}(0)$) and Eq (2):

$$X(s) = \frac{K_f V_A \omega}{(s^2 + \omega^2)(ms^2 + cs + m\omega_n^2)} + \frac{(ms + c)x(0) + m\dot{x}(0)}{ms^2 + cs + m\omega_n^2} \quad (3)$$

where $X(s)$ denotes the Laplace transform of the displacement $x(t)$ of the inner mass $m$, $k$ is the stiffness of the spring and $c$ is the constant of proportionality, the natural frequency $\omega_n$ can be determined by

$$\omega_n = \sqrt{\frac{k}{m}}. \quad (4)$$

Given the mechanical characteristics based on resonance, a resonant electromagnetic actuator is required to be driven by $u(t)$ at a resonant frequency, i.e., $\omega = \omega_n$ to maximize $|X(s)|$.

Lastly, the Q factor of the mechanical structure, which determines the sharpness of resonance, needs to be set to a certain value that enhances the vibration force of the actuator:

$$Q = \frac{1}{2\zeta} \quad (5)$$

where the damping ratio is given by

$$\zeta = \frac{c}{2\sqrt{km}}. \quad (6)$$

Empirically, in the manufacturing process, this value is set to lower than approximately 0.05 to have a high $Q$, contributing to having a strong vibration force at the same time while narrowing the operable frequency range. Once the natural frequency is shifted, it cannot help but depreciating the actuation performance.

## IV. Sensorless Drive for Resonant Frequency Tracking

In this section, we analyze electrical dynamics of the resonant electromagnetic actuator to derive the relationship between the natural frequency and the electric response of the actuation system in order to build a drive scheme that helps to estimate the back EMF, automatically track the natural frequency and produce resonance without the use of an extra physical sensor.

*A. Estimation of Back EMF and Damped Natural Frequency*

Derived from the electrical and magnetic characteristics of the coil in Fig.3, a voltage signal $u$ driven to the actuator can be expressed as

$$\begin{aligned} u(t) &= Ri(t) + \varepsilon \\ &= Ri(t) + N\frac{\Delta \Phi}{\Delta t} \end{aligned} \quad (7)$$

where $R$ is the resistance of the coil, $\varepsilon$ is the potential difference across the system, $N$ is the number of the coil turns, $\Phi$ is the combined magnetic flux produced from the coil

$$\Phi_c = \frac{Ni}{\mathfrak{R}_{TH}} \quad (8)$$

and a permanent magnet (PM)

$$\Phi_m = B_r A_r. \quad (9)$$

The remnant flux density of the magnet $B_r$ has a constant value which can be decided by the type of PM material, and the total magnetic reluctance $\mathfrak{R}_{TH}$ in Eq (8) is influenced by composition of flux paths such as the dimension of PM area $A_r$, a permeability of magnetic components and leakage fluxes over the air gap between the coil and the PM. Suppose that the displacement $x$ in Eq (3) is associated only with the linear actuation so moves in a vertical direction, $\mathfrak{R}_{TH}$ could be a function of $x$ and the flux combination $\Phi$ can be expressed in conjunction with the flux linkage $\lambda$:

$$\Phi = \Phi_c(x,i) + \Phi_m = \frac{\lambda(x,i)}{N}. \quad (10)$$

By introducing Eq (10) into Eq (7), we can obtain

$$\begin{aligned} u(t) &= Ri(t) + \frac{d\lambda(x,i)}{dt} \\ &= Ri(t) + \frac{\partial \lambda(x,i)}{\partial x}\frac{dx}{dt} + \frac{\partial \lambda(x,i)}{\partial i}\frac{di}{dt} \\ &= Ri(t) + K_b \dot{x} + L\frac{di}{dt} \end{aligned} \quad (11)$$

where the first term in the bottom of Eq (11) is a resistive voltage drop, the second term is a voltage source internally generated within the coil of the actuator, also called back electromotive force (EMF), and the last term is the inductive voltage from current variations. Here, $L$ is the inductance. Assuming that the flux linkage is linearly proportional to the displacement, the back EMF term can be assumed to have a linear relation to the velocity of the movable mass $\dot{x}$, so it can be estimated with a constant $K_b$.

Suppose that the electrical input supply to the actuation system is cut off, i.e., $u(t) = 0$, making the coil to be in high impedance, the back EMF $V_b$ can be estimated by measuring potential difference between each terminal of the coil in the actuator. This can be expressed as

$$V_b(t) = K_b \dot{x} = -L\dot{i} - Ri \qquad (12)$$

Given Eq (6) and $V_A = 0$ removing the first term in Eq (3), called the steady-state response, we can also obtain

$$X(s) = \frac{(s + 2\zeta\omega_n)x(0) + \dot{x}(0)}{s^2 + 2\zeta\omega_n s + \omega_n^2} \qquad (13)$$

which represents the transient response which depends on initial conditions. Given the initial conditions cannot be negligible during the actuation mode, Eq (13) can be expressed in the time domain as

$$x(t) = \wp_0 e^{-\zeta\omega_n t} \sin(\omega_d t + \varphi_0) \qquad (14)$$

where the magnitude constant $\wp_0$ and the phase shift $\varphi_0$ can be determined by the initial displacement $x(0)$ and the initial velocity $\dot{x}(0)$, and the periodic property of the transient response depends on the damped natural frequency $\omega_d$ which is equal to or less than $\omega_n$:

$$\omega_d = \omega_n \sqrt{1 - \zeta^2}. \qquad (15)$$

The time derivative of Eq (14) is given by

$$\begin{aligned}\dot{x}(t) &= e^{-\zeta\omega_n t}(-\zeta\omega_n \wp_0 \sin(\omega_d t + \varphi_0) + \wp_0 \omega_d \cos(\omega_d t + \varphi_0)) \\ &= \wp_d e^{-\zeta\omega_n t} \sin(\omega_d t + \varphi_d) \qquad (16)\end{aligned}$$

where $\wp_d$ and $\varphi_d$ are the magnitude and the phase delay of the velocity, respectively.

Finally, by substituting Eq (16) into Eq (12), we can derive the main result from the mathematical analysis in this section, i.e., the relationship between the back EMF and the damped natural frequency:

$$V_b(t) = K_b \wp_d e^{-\zeta\omega_n t} \sin(\omega_d t + \varphi_d). \qquad (17)$$

Given this, we can extract periodic characteristics of the actuation system, that is, the resonant frequency, through the estimation of the back EMF. As discussed above, this deduction can be valid when the moving mass is excited by a certain external voltage input before the moment when the input is stopped to supply, contributing to nonzero initial conditions.

*B. Automatic Resonant Frequency Tracking Algorithm*

Based on the relationship expressed in Eq (17), we propose a new sensorless drive scheme for resonant electromagnetic actuators through the back EMF estimation for the real-time tracking of the resonant frequency during the actuation mode. This approach does not require the use of a physical sensor. The proposed method also does not require to know, or manually check, the natural frequency preset by a manufacturer.

A unit step input can be used to oscillate the actuation system, and the induced transient response Eq (13) can be monitored through probing the tip of each input terminal of the actuator coil and estimating the back EMF. The estimated signal can be used to synchronize the next input aligned with the periodical motion of the inner mass. In the actuation model shown in Fig. 3, the velocity of the moving mass is maximized at equilibrium given that the restoring force of the elastic spring resists against the applied external force in accordance with Hooke's law [26]. Given this, our idea is to amplify the velocity at the moment when the peak of $|\dot{x}|$ is detected, by generating the external electromagnetic force as shown in Fig. 4. This helps to make the time interval between current and previous step inputs fit into the damped natural period, in turn driving the actuator at a self-sensed resonant frequency. Here, we assume that the phase shift in Eq (17) can be ignorable in reality given that the value is small during the actuation of the system designed to produce vibration feedback at the low frequency range and our focus is laid on the exploration of the periodicity.

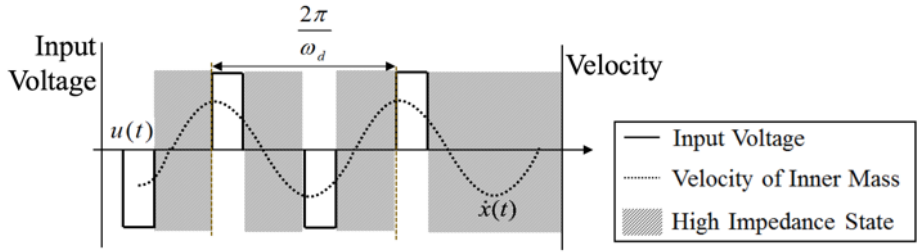

**Fig. 4.** Schematic graph for the periodic monitoring of the velocity of the inner mass of the resonant electromagnetic actuator estimated by the back electromotive force.

As can be seen in Fig. 4, iterative transitions to the high impedance state is required so as to drive the actuator and estimate the back EMF. The connection of each physical terminal of the actuator to a controller needs to be controllable for connecting it to a high impedance node for the observation of the mass movement and connecting it to an external power source for the driving. Fig. 5 describes the block diagram of the proposed sensorless drive system. A switch module in the motor driver, which is linked to *shut-down ports* of amplifiers (i.e., an audio amplifier that amplifies the current level of input sources and a voltage amplifier that increases the amplitude of the back EMF signals), helps the transition.

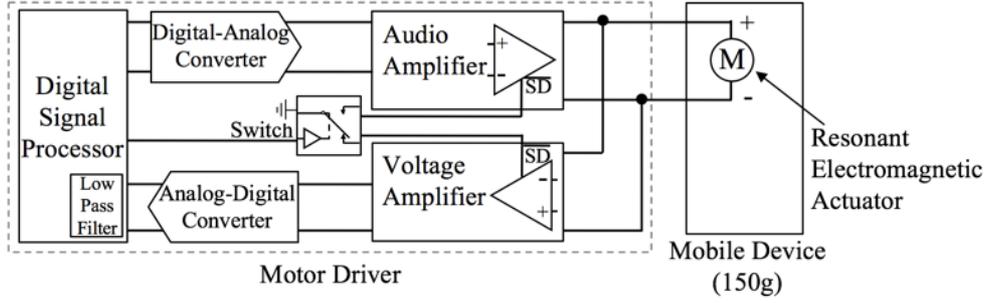

**Fig. 5.** Block diagram of the proposed sensorless resonant electromagnetic actuator drive system.

An additional application of this sensorless drive method is the residual vibration control. A vibration that lasts after the entire actuation period stops is called *residual vibration*. This affects a person's haptic perception. For the sensorless system, model-based feed-forward control methods such as the input shaping technique [15] can be used to reduce the residual oscillation. This type of techniques, however, cannot address the resonant frequency shift issue. On the other hand, the proposed approach could support the residual vibration reduction given that the system is capable of monitoring the periodical characteristics.

Algorithm 1 describes details of the proposed algorithmic flow for the automatic resonance tracking and for the reduction of residual vibrations. The differential amplified back EMF (in Fig. 5) signal can be read through `ad0_float`. This is low-pass filtered with a cutoff frequency of 500Hz, considering the sensitivity range of human mechanoreceptors [7,8] and the operable frequency range of resonance-based actuators commercialized for mobile devices. The unit step input is driven by calling a function `diff_DAC_out()` with a parameter to set a current direction within the actuator coil. The duration of each step input, `PULSE_DURATION`, is set to be equal to or less than a half period of resonant frequency $\pi/\omega_n$ for the monitoring of the mass movement during the high impedance state. A function `high_Impedance()` makes the audio amplifier (in Fig. 5) shutdown for the transition to the high impedance state. Additionally, for the residual vibration reduction, a parameter named `isBreakingTriggered` is used to invert the step input direction, and the iterative operation lasts until the filtered back EMF value becomes lower than a threshold value (e.g., 10% of the maximum value of the filtered signal).

**Algorithm 1**. Auto-tracking of resonant frequency & Reduction of residual vibration.

```
function TIMER_ISR()      ▷ being called along with the sampling frequency, fs
    ad0_float←readDiffADC()
    back_emf_y←lowpassfilter(ad0_float, 500, fs)    ▷ applying LPF with cut-off frequency 500Hz
    if driving_mode == BEMF_AND_UNIT_PULSE_I then
        if isBackEMFmonitoring then
            if (back_emf_y[0]-back_emf_y[1])<-thr &&
               (back_emf_y[1]-back_emf_y[2])>thr then    ▷ detecting a positive peak
                high_Impedance(OFF)
                if isBreakingTriggered then
                    diff_DAC_out(POSITIVE)
                else if ~isBreakingTriggered then
                    diff_DAC_out(NEGATIVE)
                end if
                isBackEMFmonitoring← false
                time_tic←0
            end if
        else if ~isBackEMFmonitoring then
            if isBreakingTriggered then
                diff_DAC_out(NEGATIVE)
            else if ~isBreakingTriggered then
                diff_DAC_out(POSITIVE)
            end if
            if ++time_tic> PULSE_DURATION then
                high_Impedance(ON)
                isBackEMFmonitoring← true
                time_tic←0
                driving_mode← BEMF_AND_UNIT_PULSE_II
            end if
        end if
    else if driving_mode == BEMF_AND_UNIT_PULSE_II then
        if isBackEMFmonitoring then
            if (back_emf_y[0]-back_emf_y[1])>thr &&
               (back_emf_y[1]-back_emf_y[2])<-thr then    ▷ detecting a negative peak
                high_Impedance(OFF)
                if isBreakingTriggered then
                    diff_DAC_out(NEGATIVE)
                else if ~isBreakingTriggered then
                    diff_DAC_out(POSITIVE)
                end if
                isBackEMFmonitoring← false
                time_tic←0
            end if
        else if ~isBackEMFmonitoring then
            if isBreakingTriggered then
                diff_DAC_out(POSITIVE)
            else if ~isBreakingTriggered then
                diff_DAC_out(NEGATIVE)
            end if
            if ++time_tic> PULSE_DURATION then
                high_Impedance(ON)
                isBackEMFmonitoring← true
                time_tic←0
                driving_mode← BEMF_AND_UNIT_PULSE_I
            end if
        end if
    end if
    if isBreakingTriggered && back_emf_y[0] < DESIRED_VALUE then
        return    ▷ To finish the residual vibration control
    end if
end function
```

## V. Experimental Validation and Results

We conducted experiments to evaluate the performance of the proposed sensorless drive. In this validation, we aim to test the accuracy in tracking the resonant frequency and the response time from the residual vibration control. Fig. 6 shows the overall experimental setup that includes the proposed sensorless driver and a resonant electromagnetic actuator. This setup was used both for collecting the frequency response profile of a resonant electromagnetic actuator using the current amplifier and driving the actuator based on the proposed scheme using the designed motor driver. An actuator sample was attached on a mock-up device which weighs 150g and has a similar dimension of smartphones. We placed this on sponge surfaces to best minimize energy loss due to the friction force and sound. The vibration force was measured through the use of an accelerometer (Bruel and Kjaer 4524) and the measured signals were amplified by a charge amplifier (Bruel and Kjaer 2692). For the conversion from the analog signal to the digital one, a data acquisition board (NI USB 6259) was used. An oscilloscope (HP 54622A) was further used to display measured signals. The sampling frequency was set to 10kHz for both the driving of the actuator and the monitoring of the back EMF and the acceleration signals.

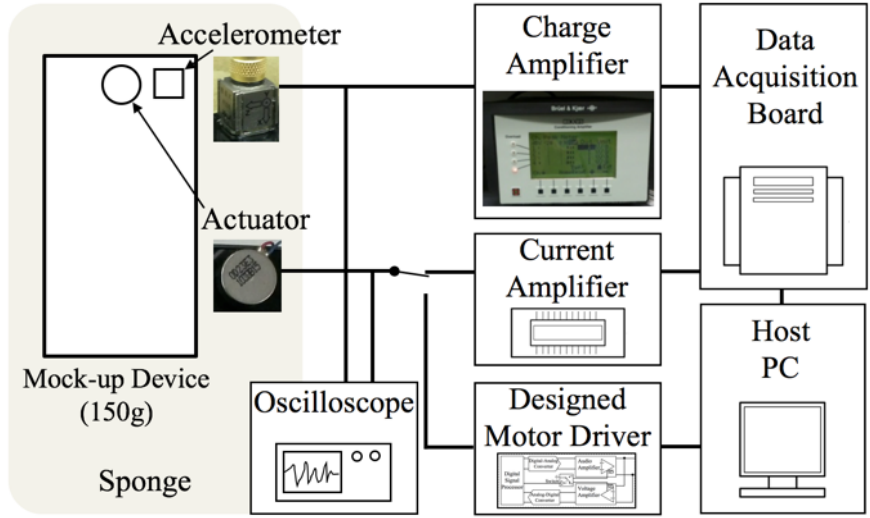

**Fig. 6.** Experimental setup to measure output accelerations of a resonant electromagnetic actuator along with the sweeping frequency and evaluate the proposed sensorless drive scheme.

*A. Automatic Tracking of Resonant Frequency*

Fig. 7 compares the system electrical output consisting of the input driving signal and the back EMF with the acceleration wave measured from the accelerometer. An actuator sample (SEMCO LRA1036, resonant frequency: 175 Hz) was driven by the proposed method for approximately 40ms. After the driving period, the actuator was made to be in the high impedance state so as to solely observe its back EMF. The top graph in Fig. 7 shows the voltage signals measured from the actuator, and the bottom graph in Fig. 7 shows the vibration acceleration wave. As in Eq (12), both measured signals had a similarity in the periodicity. Given this preliminary test result, we highlight that the proposed method is capable of driving the actuator without the requirement to know the preset natural frequency.

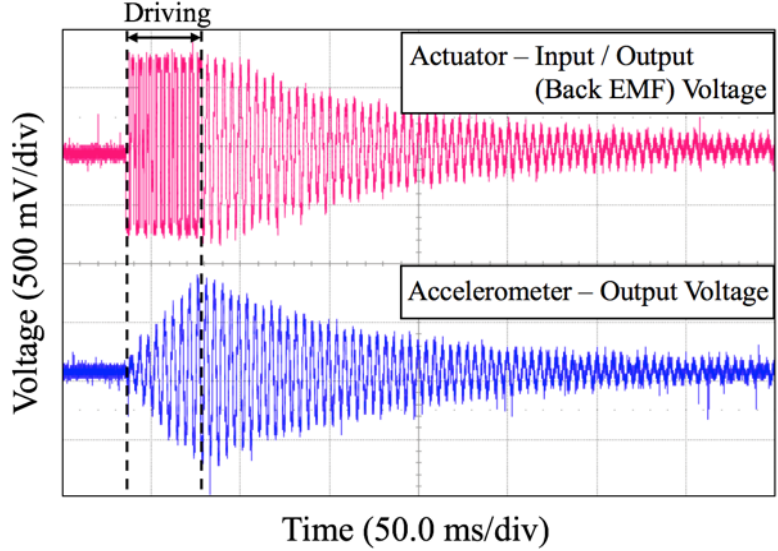

**Fig. 7.** Measured voltage (i.e., combined input voltage and output back electromotive force) between the terminals of the actuator (LRA1036, SEMCO) (top) compared with acceleration signals (bottom) while driven by our proposed method.

To validate the automatic resonance tracking accuracy, we tested two resonant electromagnetic actuators samples that have different mechanical natural frequencies (sample S1: LG LRA-SMA which has a high resonant frequency; sample S2: LG DMA which has a low resonant frequency [23]). The size dimension of both samples is 10mm x 10mm x 3mm (300 mm$^3$). Fig. 8a and 8b show the frequency response of each sample actuator which was measured through inputting sweeping sinusoidal signals of 2.5V$_{peak}$ to the actuator (i.e., (a) S1, (b) S2). Fig. 8c and 8d show automatically detected resonant frequencies of S1 and S2, respectively. Compared with the ground truth from the frequency response measurements, the errors of tracked resonant frequencies of S1 and S2 were 0.9 Hz and 0.72Hz, respectively. Given the operable frequency ranges of S1 and S2 within which the detected frequencies fall (i.e., S1: 1/(4.785ms)=209Hz, S2: 1/(6.604ms)=151.42Hz), the result demonstrates the performance of the proposed sensorless scheme in the simultaneous resonance detection and actuation. By contrast with the traditional drive scheme for resonant electromagnetic actuators that operates with a given, fixed resonant frequency (described on a datasheet provided by a manufacturer), our proposed method does not demand to know the predefined frequency and is able to adaptively cope with situations where the natural frequency shift occurs, in turn maintaining an adequate level of the vibration force.

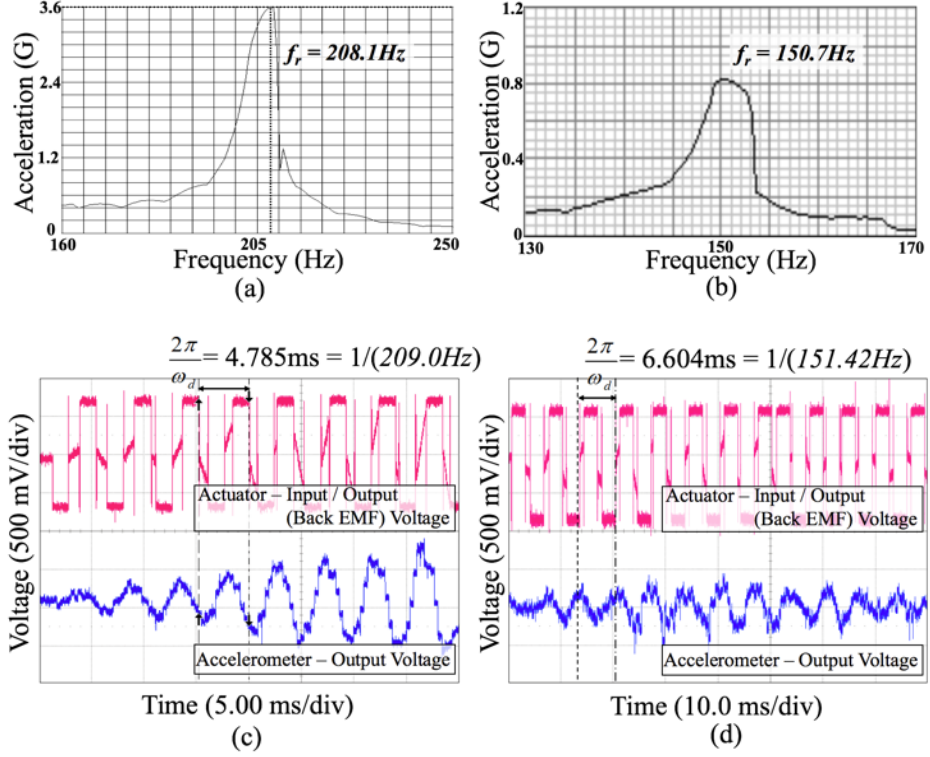

**Fig. 8.** Sensorless drives and detected damped natural frequency in accordance with the resonant electromagnetic actuators, each with a different resonant frequency (S1: LRA-SMA, LG (high mechanical resonance), LG; S2: DMA, LG (low mechanical resonance)): (a) frequency response and (c) measured input/output signals of S1, (b) frequency response and (d) measured input/output signals of S2.

*B. Residual Vibration Control*

As part of an effort to extend use cases of the proposed sensorless resonance tracking method, we conducted further experiments to test its capability in reducing the residual vibration. It is evident that residual oscillation produces differences between a designed vibrotactile pattern and a physical vibration feedback, in turn affecting a person's haptic perception. Fig. 9a shows the residual vibration produced from the actuation of the resonant electromagnetic actuator (SEMCO LRA1036) without the residual vibration control. As observed, the residual wave (> 0.9s) was yielded after the actuation period of 1.28s. From the same settings except for the parameter `isBreakingTriggered` set to `true` (see Algorithm 1), activating the residual vibration control, the duration of the residual oscillation was dropped to 0.09s as shown in Fig. 9b. It is remarkable that the proposed method reduced the stopping time of residual waves to less than 10% of the original duration. This can help to bridge the gap between the designed and actual vibration patterns.

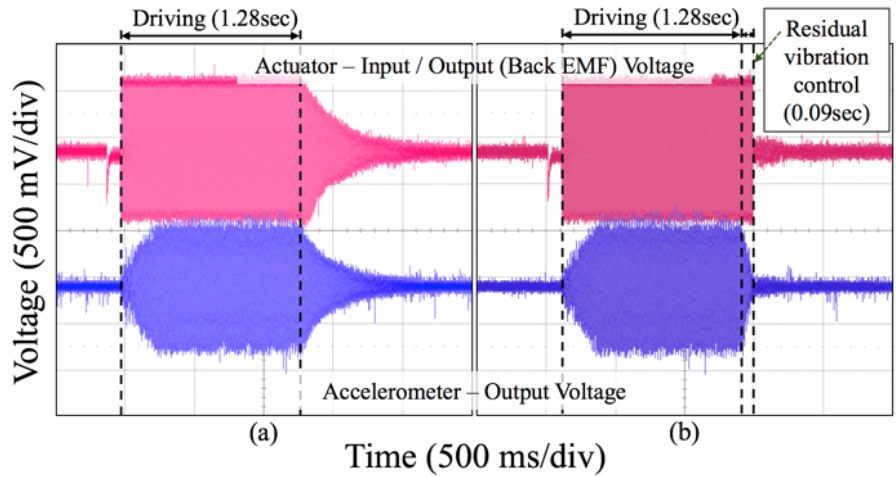

**Fig. 9.** Measured voltage between the terminals of the actuator (LRA1036, SEMCO) compared with acceleration signals while driven by our proposed method: (a) without, (b) with the proposed residual vibration control.

To highlight an advantage of the residual vibration control, we applied the method to the creation of short haptic feedback for a virtual button to mimic the feedback from a physical button, one of interesting topics in tactile interaction for mobile devices (e.g., [27]). Fig. 10 compares the vibration pattern produced from the residual vibration control based on the sensorless resonance tracking (left in Fig. 10) with the pattern from the traditional method (right in Fig. 10). The duration of each actuation input signal driven to the actuator (SEMCO LRA1036) was approximately 30ms. The proposed method was capable of controlling the residual vibration, quickly dropping the vibration amplitude lower than the steady-state error margin (10% of the maximum acceleration, here, approximately 0.05G). Compared to the traditional actuation method without the residual vibration control, the proposed approach reduced the residual duration from 200ms to 65ms (by 67.5%).

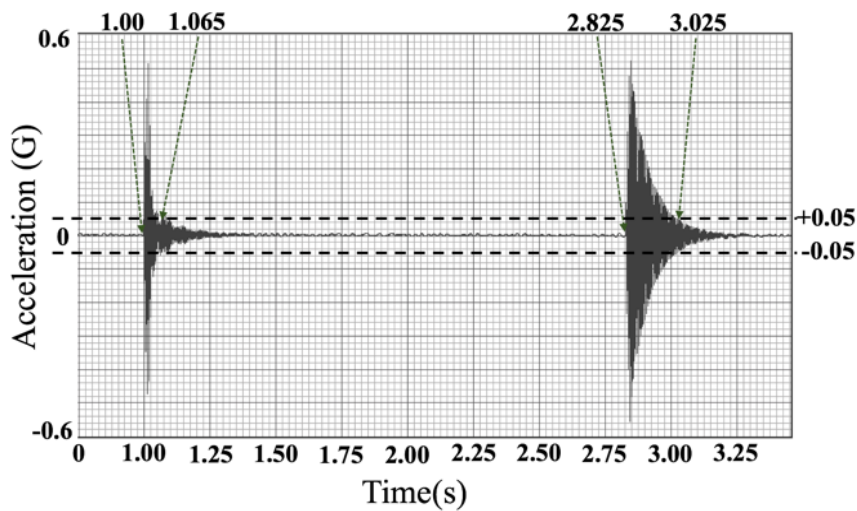

**Fig. 10.** Creation of short vibration feedback for a virtual button in mobile devices - conveying '*button click*' sensation: the proposed method with the residual vibration control (left), the conventional drive (i.e., using a sine wave input at a fixed resonant frequency) (right).

## VI. Conclusion

In this paper, we have proposed a new sensorless drive technique to automatically track the resonant frequency of a resonant electromagnetic actuator which is designed for mobile devices. The method is grounded in the mathematical analysis of the actuation model, which derives the relation of the back electromotive force to the damped natural frequency of the moving mass inside an actuator. The developed drive system and algorithm are to drive the actuator and automatically track its unknown damped natural frequency through the back EMF estimation so as to generate vibration feedback. This does not require to use additional sensors. Conducted experiments have demonstrated the robustness of the proposed approach in automatic resonance tracking with a good accuracy result (lower than 0.5% errors). Given this, it is expected that the method can improve an actuator's lifetime by adaptively addressing the resonant frequency shift issue, and potentially increase yield rates during manufacturing processes (for example, decreasing the failure rate from drop tests [28]), helping to reduce production costs. Lastly, we have demonstrated the capability of the method in the residual vibration control for the creation of short and strong vibrotactile effects. We believe that the ability can help to create rich tactile feedback that can assist a person's cognitive process (e.g., pen writing [29]) or emotional self-regulation [30,31].

# Supporting Information

**Figure legends**

**Fig. 1.** Schematic mechanical structure of the resonant electromagnetic actuator.

**Fig. 2.** Measured the shifted resonant frequency of the resonant electromagnetic actuator (Product: LRA – SEMCO 1036) along with the different peak voltage of input signal on the frequency response: (a) 175.5Hz - 1.0$V_{peak}$, (b)174.1Hz - 1.2$V_{peak}$, (c)173.4Hz - 1.5$V_{peak}$.

**Fig. 3.** Combined electrical circuit and mechanical model of the actuator embedded in mobile devices.

**Fig. 4.** Schematic graph for the periodic monitoring of the velocity of the inner mass of the resonant electromagnetic actuator estimated by the back electromotive force.

**Fig. 5.** Block diagram of the proposed sensorless resonant electromagnetic actuator drive system.

**Fig. 6.** Experimental setup to measure output accelerations of a resonant electromagnetic actuator along with the sweeping frequency and evaluate the proposed sensorless drive scheme.

**Fig. 7.** Measured voltage (i.e., combined input voltage and output back electromotive force) between the terminals of the actuator (LRA1036, SEMCO) (top) compared with acceleration signals (bottom) while driven by our proposed method.

**Fig. 8.** Sensorless drives and detected damped natural frequency in accordance with the resonant electromagnetic actuators, each with a different resonant frequency (S1: LRA-SMA, LG (high mechanical resonance), LG; S2: DMA, LG (low mechanical resonance)): (a) frequency response and (c) measured input/output signals of S1, (b) frequency response and (d) measured input/output signals of S2.

**Fig. 9.** Measured voltage between the terminals of the actuator (LRA1036, SEMCO) compared with acceleration signals while driven by our proposed method: (a) without, (b) with the proposed residual vibration control.

**Fig. 10.** Creation of short vibration feedback for a virtual button in mobile devices - conveying '*button click*' sensation: the proposed method with the residual vibration control (left), the conventional drive (i.e., using a sine wave input at a fixed resonant frequency) (right).

**Algorithm legends**
**Algorithm 1**. Auto-tracking of resonant frequency & Reduction of residual vibration.